
\magnification=\magstep1  \overfullrule=0pt
\advance\hoffset by 0.65truecm
\def\lb{\lbrack}\def\rb{\rbrack}  \def\q#1{$\lb#1\rb$}
 \def\np{{\rm Nucl.\ Phys.\ B\ }}

\def\intj{{\rm Int.\ J.\ Mod.\ Phys.\ A\ }}
\def\bn{\bigskip\noindent} \def\mn{\medskip\smallskip\noindent}
\def\sn{\smallskip\noindent} \def\crns{\cr\noalign{\smallskip}}
\def\l{\hbox{$\cal L$}}  
\def\kk{\lb k/2\rb} \def\dil{\hbox{dilogarithm}}
\def\mod{{\rm mod}}  
\def\L{\hbox{\rm L}} \def\Li{\hbox{\rm Li}}
\def\log{\hbox{\rm log}} \def\cef{c_{{\rm eff}}}
\def\ie{i.e.\ } \def\eg{e.g.\ }  \def\asy{asymptotic}

\def\Z{\hbox{{\rm Z{\hbox to 3pt{\hss\rm Z}}}}}
\def\zp{\hbox{${\rm Z{\hbox to 3pt{\hss\rm Z}}}_+$}}
 \font\extras=cmss10 scaled 750
\setbox2=\hbox{{{\extras Z}}}
\setbox3=\hbox{{{\extras z}}}

\def\C{{\rm C \kern-5.5pt I \ }}
\def\Q{{\rm Q \kern-5.5pt I \ }}
\def\R{\hbox{{\rm I{\hbox to 5.5pt{\hss\rm R}}}}}
\font\grosss=cmr7 scaled \magstep4

{\nopagenumbers \null  \vfill
\centerline{\grosss Dilogarithm Identities in Conformal Field Theory}
\bigskip\bigskip
\centerline{Werner Nahm, Andreas Recknagel, Michael Terhoeven}
\bigskip
\centerline{Physikalisches Institut der Universit\"at Bonn}
\centerline{Nussallee 12, D-53oo Bonn 1}\centerline{Germany}
\vfill  \centerline{\bf Abstract} \sn
Dilogarithm identities for the central charges and conformal
dimensions exist for at least large classes of rational
conformally invariant quantum field theories in two dimensions.
In many cases, proofs are not yet known but the numerical and
structural evidence is convincing. In particular, close relations
exist to fusion rules and partition identities. We describe some
examples and ideas, and present conjectures useful for the
classification of conformal theories. The mathematical structures
seem to be dual to Thurston's program for the classification
of 3-manifolds. \bn\vfill
\line{e-mail: UNP01A or UNP044 at ibm.rhrz.uni-bonn.de \hfill
November 1992}
\eject}
\pageno=1 \leftline{\bf 1. Introduction} \bn
Recently, the Rogers \dil\ function has appeared in several places
in physics. This function (cf. \q{20}), defined by
$$\eqalign{&\L(z) = \Li_2(z) + {1\over2}\log(z)\log(1-z), \cr
           &\Li_2(z) = \sum_{i=1}^{\infty}{z^n\over n^2} =
                   - \int_0^z {\log(1-w)\over w} dw \cr} \eqno(1.1)$$
for $0\leq z\leq 1\,$, is also characterized as the
unique function that is three times differentiable and satisfies
$\L(1)={\pi^2\over6}$ as well as the "five term relation"
$$ \L(w)+\L(z) + \L(1-wz) +  \L\Bigl({1-w\over1-wz}\Bigr) +
\L\Bigl({1-z\over1-wz}\Bigr) = 3\L(1)\,, \eqno(1.2) $$
from which simpler functional equations can be derived, \eg
$$ \eqalign{&\L(1-z) = \L(1) - \L(z)\,, \cr
 &\L(z^2) = 2\L(z)- 2\L\Bigl({z\over 1+z}\Bigr)\,. \cr} \eqno(1.3) $$
$\L(z)$ may be extended consistently to real $z>1$ with the help of
$$ \L(z) = 2\L(1) - \L(1/z)\ . \eqno(1.4) $$    Beyond
these relations, the Rogers \dil\ has other intriguing properties,
which make it a very interesting function for various branches
of mathematics such as number theory including algebraic $K$-theory,
geometry  of hyperbolic 3-manifolds, and even Grothendieck's theory of
motives. \q{26,3,8}          \sn
In physics, the \dil\ has shown up in the context of integrable
2-dimensional quantum field theories and lattice models
\q{2,15,16,23}. More precisely, when
studying the UV limit or the critical behaviour of such systems
-- typically by methods involving the Thermodynamic Bethe Ansatz (TBA)
\q{15,27} -- it was observed that \eg the central charge of the
"associated" conformal field theory can be expressed through the
\dil\ evaluated at certain algebraic numbers. One aim of
this article is to show that these relations already emerge within
a pure CFT context. Some of the results have been presented
previously \q{21}.           \sn
In the next section it is demonstrated how \dil\ identities
can be derived from the asymptotics of character identities of
Rogers-Ramanujan type, mainly
concentrating on the examples of non-unitary $c(2,q)$ Virasoro minimal
models. However, the method used, which is a slight modification of the
one outlined in \q{24}, is applicable to more general cases
and should allow to prove some of the various conjectures
on \dil\ identities stated previously in the literature.
Section 3 contains some formulas connecting the central
charge and the conformal dimensions of a RCFT to the fusion rules
by means of the Rogers \dil. The fusion rules here seem to replace
the functional relations from the TBA
encountered in the above mentioned applications. We conclude
with some further conjectures on possible generalizations and give
examples of their power in view of the CFT classification problem.
In addition, we sketch the geometric background of the \dil, which
could well prove to be the most important feature about its
appearance in CFT.   \bn\bn
\leftline{\bf 2. Dilogarithm Identities from Character Asymptotics}\bn
We consider the non-unitary Virasoro minimal models with central charge
$$  c(2,k+2)= 1-{3k^2\over k+2} \eqno(2.1) $$
($k\geq 3$, odd) and primary fields of conformal dimensions
$$h_j=-{j(k-j)\over2(k+2)},\quad j\in\{0,1,\ldots,\lb k/2\rb\}.
\eqno(2.2)$$
$h_j$ is minimal for $j=\lb k/2\rb$, and the effective central
charge $\cef=c-24h_{{\rm min}}$ thus takes the value
$$\cef={k-1\over k+2} .\eqno(2.3) $$
The \dil\ identities will arise from studying the \asy\ behaviour
of the character functions
$$ \chi_{j}^{}(\tau) = q^{h_j-c/24}
   \prod_{n\not\equiv 0,\pm(j+1) \mod (k+2)} (1-q^n)^{-1}\eqno(2.4) $$
where $q=e^{2\pi i\tau}$; they can also be written in sum-form \q{1,4}
$$ \prod_{n\not\equiv 0,\pm(j+1) \mod (k+2)} (1-q^n)^{-1}
     = \sum_{n_1,\ldots,n_{\kk}\geq 0}
     {q^{N_1^2+\ldots +N_{\kk}^2+N_{j+1}+\ldots +N_{\kk} }\over
      (q)_{n_1}\ldots (q)_{n_{\kk}} },   \eqno(2.5)  $$
where $k$ and $j$ satisfy the same constraints as above,
$N_i=n_i+ \ldots +n_{\kk}$
for $i\in\{ 1,\ldots,\kk\}\,$, and $(q)_n=(1-q)\cdots(1-q^n)\,$.
(2.5) are generalized Rogers-Ramanujan identities, also known as
Andrews-Gordon identities.  \mn     In the following,
we study the \asy\ growth of the character functions. On the one
hand, this follows quite abstractly from the modular
properties of RCFT characters: Elaborating arguments given in
\q{5} (cf. also \q{9}), one sees that in the limit
$t=-i\tau\rightarrow 0^+ $ (\ie $q\rightarrow 1^-$)
$$  \chi_{j}^{}(\tau)=\sum_{l}S_{j,l}\,\chi_{l}^{}\Bigl({1\over-\tau}
\Bigr) =S_{j,\min}\,e^{\pi\cef /12t}
   \Bigl\lb 1 + \sum_{k\neq\min}{S_{j,k}\over S_{j,\min}}
\,e^{-2\pi(h_{k}-h_{\min})/t} +\ldots\Bigr\rb \,, \eqno(2.6) $$
where $S$ represents the modular transformation $\tau\rightarrow
-1/\tau$ on the character functions, in case of the $c(2,k+2)$ models
given by
$$  S_{i,j}=\sqrt{4\over k+2} (-1)^{i+j +(k-1)/2}
    \sin\Bigl( {2\pi(i+1)(j+1)\over k+2} \Bigr)\ . \eqno(2.7) $$
Note that above quite naturally "generalized quantum dimensions"
${\cal D}^{(i)}_j = {S_{j,i}\over S_{j,min}} $ appear.    \sn
On the other hand, the \asy s of the coefficient $a_M$ in
$$\chi_j^{}(\tau)=q^{h_j-c/24}\sum_{M=0}^{\infty}a_M q^M \eqno(2.8)$$
can be calculated concretely from the sum-form, and if $a_M$ diverges
\eg as $\exp\{2\sqrt{M\l}\}$ (which is just what we will find below),
we can approximate
$$\eqalign{ \sum_M a_M q^M &\simeq \int\!dM\,a_M q^M
= 2\exp\{\l/2\pi t\} \,
   \int_0^\infty \!dx\,x \exp\{-2\pi t(x-\sqrt{\l}/2\pi t)^2\} \cr
      &\sim \exp\{\l/2\pi t\}\,,\cr}  \eqno(2.9)  $$
which by comparison with (2.6) leads to an equation for the effective
central charge
$$ \cef = {6\over\pi^2} \l\,.  \eqno(2.10) $$      \sn
For the calculation of $\l$ we start from the sum-side of the
Andrews-Gordon identities (2.5). In particular, since the leading
order of the \asy\ growth is the same for all characters of a given
$c(2,2+k)$ model, we are free to choose the one of the $h_{{\rm min}}=
h_{\lb{k/2}\rb}$ primary field
$$ \sum_{M=0}^{\infty} a_M q^M
     = \sum_{n_1,\ldots,n_{\lb{k/2}\rb}\geq 0}
  {q^{{\bf n}B{\bf n}^t}\over (q)_{n_1}\cdots(q)_{n_{\lb{k/2}\rb}}}\ ,
  \eqno(2.11)$$
where ${\bf n}=(n_1,\ldots,n_{\lb k/2\rb})\,$, and $B$ is the
inverse of the Cartan matrix of the tadpole graph $A_{k+1}/\Z_2\,$.
\sn
By Cauchy's theorem, $a_{M-1}$ can be expressed as an integral
$$ a_{M-1} = \oint {dq\over2\pi i} \sum_{{\bf n}\geq0}
{q^{{\bf n}B{\bf n}^t-M}\over (q)_{n_1}\cdots(q)_{n_{\lb{k/2}\rb}}}\ .
\eqno(2.12) $$
The ideas how to obtain the large $M$ limit of (2.12) are
given in \q{24}: The keyword is "saddle point approximation", meaning
that a crude estimate of the integral can be obtained from
the integrand evaluated at its saddle point.
In the following we present a simplified version of the
procedure, which yields the correct leading term. For more details
and for error estimates using the Euler-Maclaurin formula
the reader is referred to \q{24}.        \sn
First as in (2.9) we replace the
summation by an integration over $dn^{\lb k/2\rb}$, treating the
$n_i\,$'s as continuous variables. Finding the saddle point amounts
to setting the partial (logarithmic) derivatives of the integrand
$f(q,{\bf n})$ to zero. Before doing so, we use Euler-Maclaurin a
second time for the denominator
$$ \log(q)_{n_i} \simeq \int_0^{n_i}\!dk\,\log(1-q^k) \eqno(2.13)$$
so that the logarithm of the integrand is approximately given by
$$ \log f(q,{\bf n}) \simeq ({\bf n}B{\bf n}^t -M)\,\log q
 -\sum_{i=1}^{\lb k/2\rb} \int_0^{n_i}\!dk\,\log(1-q^k)\ .\eqno(2.14)$$
Now the saddle point conditions $\partial_{n_i} \log f =0$ give
a set of algebraic equations on $\xi_i := q^{n_i}$
$$ 1-\xi_i  = \prod_{j=1}^{\kk} \xi_j^{(B_{ij}+B_{ji})}\ . \eqno(2.15)$$
Note by the way that (2.15) is a
parameter-independent variant of equations occuring in the
TBA technique \q{15,27,23}. Moreover, as we will see below,
the $\xi_i$ are closely related to
the quantum dimensions ${\cal D}_i$ of the representation with
highest weight $h_i$, defined as the ratio of the $a_M$ for large $M$,
or \q{5,7}
$$ {\cal D}_i = \lim_{q\to1^-}{\chi_i(q)\over\chi_0(q)} = {S_{i,0}
\over S_{0,0}}\ .  \eqno(2.16) $$
Before differentiating (2.14) with respect to $q$ we plug these
equations into $\log\, f(q,{\bf n})$ and obtain -- after having used
several substitutions and the definition (1.1) of the Rogers \dil\ --
$$\int_0^{n_i}\!dk\,\log(1-q^k)={n_i\over\log\xi_i}\Bigl\lb\L(1-\xi_i)
+{1\over2}\log(1-\xi_i)\,\log(\xi_i)\,\Bigr\rb \eqno(2.17) $$
and furthermore
$$ \log f(q,{\bf n}) = -M \log q - {1\over \log q}
\sum_{i=1}^{\kk} \L(\delta_i)\,.  \eqno(2.18)$$
Here $\delta_i = 1-\xi_i$ has been introduced for convenience.
Now $\partial_{q} \log f =0$ fixes $q$ at the saddle point
$$ (\log q)^2 = {1\over M} \sum_{i=1}^{\kk}\L(\delta_i)\ ,\eqno(2.19)$$
so that finally the \asy\ behaviour of $a_M$ is given by
$$a_M\simeq f(q,{\bf n})|_{s.p.}=\exp\{2\bigl(M\sum\L(\delta_i)\bigr)^{
1\over2}\,\}\ . \eqno(2.20)$$    Together with (2.10) this allows
us to express the effective central charge in terms of the \dil.
$$ \cef = {1\over\L(1)} \sum_{i=1}^{\kk}\L(\delta_i)\ . \eqno(2.21)$$
It remains to solve the algebraic equations (2.15), \ie to determine
$\delta_i$ in
$$\delta_i = \prod_{j=1}^{\kk}(1-\delta_j)^{2B_{ij}}\ .\eqno(2.22)$$
After some manipulations, using $B=(2-I)^{-1}$, where $I$ denotes the
adjacency matrix of the graph $A_{k+1}/\Z_2\,$, and with the
substitution $\delta_i = 1/d_i^2\,$, (2.22) can be brought into
the more familiar form
$$d_i^2=1 + \prod_{j=1}^{\kk}d_j^{I_{ij}}=
\cases{ 1+d_2 &$i=1\,$, \cr
       1+d_{i-1}d_{i+1} &$i=2,\ldots,\kk-1\,$, \cr
       1+d_{\kk-1}d_{\kk} &$i=\kk\,$, \cr} \eqno(2.23)$$
which is just one way to recursively define the
\v Cebyshev polynomials of the second kind, together with
an additional "truncation condition"; put differently,
$$d_i={\sin{(i+1)\pi\over k+2}\over\sin{\pi\over k+2}} \eqno(2.24)$$
solves (2.20).
Rephrasing (2.21), we have derived formulas telling that the values
of the \dil\  at certain algebraic numbers (here:\ Jones indices)
sum up to a rational result (except for a factor $\L(1)$)
$$ \sum_{i=1}^{\kk}
 \L\Bigl({\sin^2{\pi\over k+2}\over\sin^2{(i+1)\pi\over k+2}}\Bigr)
= {\pi^2\over6} {k-1\over k+2}\ . \eqno(2.25)$$
Of course, this method to extract \dil\ identities from the
\asy s of the character functions can also be applied to other
cases -- provided a sum form of the characters similar to (2.5) is
known \q{4,12}.         \sn
Furthermore, in view of the next-to-leading term in the expansion
(2.6), it should be possible to obtain \dil\ expressions for
$c-24h_i$ as well, namely from the \asy\ behaviour
of suitable linear combinations of characters.
\bn\bn{\bf 3. On the Dilogarithm and Fusion Rules}\bn
There seems, however, to be another way to compute conformal dimensions
with the help of the Rogers \dil. To see this, let us once more
consider formula (2.25). This equation is already well known in the
literature, and was derived previously by Kirillov \q{13} from
the functional relations of $\L(z)$. As was stated by this author,
the r.h.s.\ of (2.25) is also related to the central charge of the
SU(2) level $k$ WZW model. On the other hand,
the arguments of the \dil\ on the l.h.s.\ are just the inverse
squares of the quantum dimensions of the primary fields of this WZW
theory (or the $c(2,k+2)$-model), which can alternatively be
characterized as the unique maximal eigenvalues $\lambda_1(i)$
of the fusion matrices $N_i$ associated to the primary fields.
This observation suggests to replace the quantum dimensions
by other eigenvalues $\lambda_j(i)$ of the $N_i\,$;
then the following seems to hold
$${1\over\L(1)}\,\sum_{i=2}^{k+1}\L\Bigl({1\over\lambda_j^2(i)}\Bigr)
=  c_k - 24h^{(k)}_{j-1} +6(j-1)\ , \eqno(3.1)$$
where  $$  c_k = {3k\over k+2}\ \ {\rm and}\ \
 h_{j-1}^{(k)} = {j^2-1\over4(k+2)}\ , j=1,\ldots,k+1 \eqno(3.2) $$
are the central charge and the
conformal dimensions of the SU(2) WZW primary fields, respectively,
and the eigenvalues of the fusion matrices -- closely related
to the $S$-matrix or the "generalized quantum dimensions" mentioned
before -- are given by
$$ \lambda_j(i) = {S_{ij}\over S_{i1}}=
{\sin{ij\pi\over k+2}\over\sin{j\pi\over k+2}}\
, \ \ i,j = 1,\ldots,k+1\,. \eqno(3.3) $$
We have checked (3.1) explicitly for low values of the level, and
proved it for some special cases (\eg for
$j=2$, $k$ arbitrary) along the lines of Kirillov.
An analogous formula holds for the $c(2,k+2)$ models, with
$c$ replaced by $\cef$ and $h_j$ by $h_j-h_{{\rm min}}\,$.
\sn
In the literature, formulas for other WZW models based on a
Lie algebra $X$ of rank $r$ have appeared. To this end,
one defines functions $Q_m^{(a)}(z)$ on the complexified
weight lattice of $X$, which are subject to the following functional
relations (for simplicity, we only deal with simply laced $X$ in
the following, for the general case see \q{13,14,18})
$$ 1- {Q_{m-1}^{(a)}Q_{m+1}^{(a)}\over Q_{m}^{(a)}{}^2} =
\prod_{b=1}^r Q_{m}^{(b)}{}^{-C^X_{ba}} \eqno(3.4) $$
(here $C^X$ is the Cartan matrix of $X$) and satisfy the initial
conditions $Q_{-1}^{(a)}=0$, $Q_{0}^{(a)}=1$ for $a=1,\ldots,r$.
The functions $Q_{m}^{(a)}$ have been first introduced in \q{14}
in the study of Lie group representations by means of Yangians,
where it can be seen that they are essentially built up from
classical Lie group characters. Using these quantities,
Kirillov conjectured a \dil\ formula for the central charge of the
$X$ WZW model \q{13}: Denoting the l.h.s.\ of (3.4) by
$f_m^{(a)}(z)$ and the central charge of the level $l$ theory by
$c^X_l$, Kirillov claims that
$$\sum_{a=1}^r\sum_{m=1}^{l-1}\L(f_m^{(a)}(0))
={\pi^2\over6}(c^X_l-r)\,.\eqno(3.5)$$
He proved this formula for the $X=A$ and $X=D$ cases.
In \q{17,18}, conjecture (3.5) has been generalized
to non-simply laced Lie algebras and also, more important, to
formulas involving other entries of the $f_m^{(a)}(z)$:
Taking $z=\Lambda$ to be a dominant highest weight satisfying
the usual level restriction, these authors have claimed that
via the \dil\ also the conformal dimensions $h^X_l(\Lambda)$
of the Kac-Moody representation labelled by $\Lambda$
can be determined -- if some correction terms are introduced
$${1\over\L(1)}\sum_{a,m}\tilde{\L}(f_m^{(a)}(\Lambda))=
(c^X_l-r -24\tilde h^X_l(\Lambda))\ (\mod\ 24)\,, \eqno(3.6)$$
where the sums have the same ranges as in (3.5).
$\tilde{\L}(f_m^{(a)})$ is given by (with $\arg(z)\in(-\pi,\pi\rb\,$)
$$\eqalign{\tilde{\L}(f_m^{(a)}) &= \Li_2(f_m^{(a)}) +{1\over2}
\log|f_m^{(a)}|\ \log|1-f_m^{(a)}| \cr
&\phantom{xx}+{1\over2}\sum_{b,k}\arg(1-f_m^{(a)})
C^X_{ab}(C^{(l-1)}_{mk})^{-1}\arg(1-f_k^{(b)}) \cr}\eqno(3.7) $$
- where $C^{(l-1)}$ is the Cartan matrix of $A_{l-1}$ -
and the corrected highest weights are
$$ \tilde h^X_l = h^X_l-\sum_{a,b=1}^r a^{(a)}C^X_{ab}a^{(b)}
\eqno(3.8a)$$        with
$$ a^{(a)} = l\arg Q_{l-1}^{(a)}-l\arg Q_{l}^{(a)} +
\sum_{j=1}^{l-1} j\arg(1-f_{j}^{(a)})\,. \eqno(3.8b)$$
All the functions $f_m^{(a)}$ and $Q_{m}^{(a)}$ in (3.7,8) are
evaluated at the highest weight $\Lambda$.
Then, according to \q{18}, "numerical checks indicate" that the
$Q_{m-1}^{(a)}$ in addition to the functional relation (3.4) satisfy
also $$ Q_{m}^{(a)} = Q_{l}^{(a)}Q_{l-m}^{(a)\, *}, \eqno(3.9a)$$
which in particular enforces a truncation
$$Q_{l+1}^{(a)} =0\ . \eqno(3.9b) $$
In \q{18} the quantities $Q_m^{(a)}$ were regarded as arising from a
TBA connected to the Lie algebra $X$. However, they can also be
interpreted completely within the CFT framework, namely in terms of
the $X$ WZW model fusion rules. Considering the simplest case $X=A_1$
first, one easily sees that the functional relations (3.4) together
with (3.9) are just the SU(2) level $l$ fusion rules. For $X=A_r$,
the $Q_m^{(a)}$ subject to (3.4,9) still can be identified with
part of the WZW fusion generators, if the complex conjugation
in (3.9) is translated into conjugation of the sector. In these
cases, evaluating the functions $Q_m^{(a)}$ at various highest
weights corresponds to replacing the formal fusion ring generators
by their various eigenvalues, which can be expressed
through the $S$-matrix as 'generalized quantum dimensions'
$$Q_m^{(a)}(\Lambda) = {S_{m\Lambda_a;\Lambda}\over S_{0;\Lambda}},
\eqno(3.10)$$  where $\Lambda_a$ are the fundamental weights of $X$.
For the other algebras $X$, the situation is more
complicated, but still the $Q_m^{(a)}(\Lambda)$ can be expressed as
linear combination of such S-matrix ratios, \ie of fusion matrix
eigenvalues, with positive integer coefficients. This can be seen
from the original definition of the $Q_m^{(a)}$ in terms of
classical Lie algebra characters \q{14,17,18}, which in turn give
the WZW S-matrix elements -- see \eg \q{9}, chapter 13.
{}From all this we can conclude that the $f_m^{(a)}$ in the \dil\
can be expressed as rational functions of the fusion ring generators,
and thus within CFT the fusion rules are the algebraic structure
underlying the \dil\ identities (3.1,6).  \sn
This re-interpretation allows to stay
within CFT, but unfortunately it gives no explanation of the
argument correction terms entering the \dil\ conjecture above.
Note, however, that all these terms vanish if the entries $f_m^{(a)}$
are within the interval $(0,1)$ -- which is the case
for the original formulas given by Kirillov, $\Lambda=0$:
Then the $Q_m^{(a)}(0)$ are linear combinations of quantum dimensions
of WZW primary fields so that $Q_m^{(a)}(0) \geq1$ and
due to (3.4) we have $0\leq f_m^{(a)}(0)\leq1\,$. \hfil\break
\noindent The arg-corrections in $\tilde{\L}(z)$
at least appear understandable if one looks at the functional
equations of the $f_m^{(a)}$, which are determined by the same
matrix $C^X\otimes(C^{(l-1)})^{-1}$, see (3.15,17) below.
In the last section, we will say a little more on their possible
background. On the other hand, the origin
and the explicit form of the corrections in the
conformal dimensions $\tilde h^X_l$ are more mysterious.
According to \q{18} the $\tilde h^X_l$ coincide
with a subclass (or including corrections depending
on the path of analytic continuation with all)
of the conformal dimensions of the corresponding
parafermionic theory. However, from the CFT point of view it is
unclear why the parafermion theory shows up, since we use the
fusion rules of the full WZW theory to generate the arguments of
the \dil.       \sn
Note that the conjecture (3.1) given for SU(2) is not just a
special case of (3.6), which means that one can hope to find  -- for
each $X$ separately -- a suitable modification of the \dil\ leading to
simpler \dil\ formulas for the conformal dimensions. In other
words: There might be a special "\dil\ function" for each
fusion ring yielding "rational invariants" of this algebra in the form
of central charges and conformal dimensions. Therefore we
extend the above conjectures to the following:
\smallskip
\item{} For any RCFT with central charge $c$ and conformal dimensions
$h_j$ there are rational functions
$f_i(N_1, \ldots , N_n)$ in the generators of the fusion ring such
that if we insert the eigenvalues $\lambda_j(i)$ for $N_i$,
$${1\over\L(1)}\,\sum_i\tilde{\L}(f_i(\lambda_j(1),\ldots,\lambda_j(n)))
     = c-24h_j+ R_j\,, \eqno(3.11)$$
holds with $R_j$ an integer remainder. $\tilde{\L}$ is a modified
\dil\ function which for real arguments coincides with the Rogers \dil.
\sn       There is an algebraic constraint
on the rational functions $f_i$ which can appear as arguments of the
modified \dil\ in (3.11). In order to formulate it, we first have to
introduce the $\Z$-module $\Lambda^2(F^{\times})$, the second
exterior power of the multiplicative group of a field $F$.
($F^{\times}=F-\{0\}$ is a vector space over $F$
with the field multiplication
as "summation operation". For our applications, $F$ can be $\C$ or
the quotient field of the fusion ring of some RCFT.) In
$\Lambda^2(F^{\times})$, the following rules hold (with
$x,y,z,1\in F^{\times}\,$)
$$x\land y= -(y\land x)\,,\ \ (xy)\land z=x\land z+y\land z\,,\ \
  (\pm1)\land z = 0\,. \eqno(3.12)$$
Now consider the map $\beta:F^{\times}\rightarrow\Lambda^2(F^{\times})$
 defined by  $$ \beta(z) = z \land (1-z)\,. \eqno(3.13) $$
That this map is relevant for the \dil\ can be expected from the
differential
$$d\L(z)={1\over2}(\log(z)d\log(1-z)-\log(1-z)d\log(z))\,.\eqno(3.14)$$
$\beta(z)$ algebraically encodes the antisymmetry of $d\L(z)$ as
well as the functional properties of the logarithms in (3.14) --
compare (3.12). Beyond that, it also plays a certain role in geometry
of hyperbolic 3-manifolds, as well as in algebraic $K$-theory
(especially for the so-called Bloch group, see \q{26}).
\hfil \break
\noindent For our purposes, this construction is
interesting because of the following theorem, which was pointed
out to us by A.\ Goncharov:\smallskip
\item{} If for a finite set of algebraic numbers $\{x_i\}$ one has
$\sum_i \L(x_i) = {\pi^2\over6} q$ with $q$ rational, then
$\sum_i \beta(x_i) =0$ in $\Lambda^2(F^{\times})\,$. \sn
Because eigenvalues of fusion matrices always are algebraic numbers,
so are the arguments $f_i$ of the \dil\ in (3.11), and thus
the condition $\sum \beta(f_i)=0$ might lead, upon
thorough investigation of its consequences, to a definition of the
rational functions $f_i$ for arbitrary RCFT fusions rings. \hfil\break
\noindent Indeed one easily sees that the "$\beta$-condition" holds
for all the concrete cases listed above: Whenever the entries $f_i$
satisfy Bethe-Ansatz-like equations
$$ f_i = \prod_j (1-f_j)^{\hat B_{ij}} \eqno(3.15)$$
with a symmetric matrix $\hat B\,$, then
$$ \sum_i \beta(f_i) =  \sum_i f_i \land (1-f_i) = \sum_i
\prod_j (1-f_j)^{\hat B_{ij}} \land (1-f_i)
= \sum_{ij} \hat B_{ij} (1-f_i) \land (1-f_j) = 0 \eqno(3.16) $$
by (anti-)symmetry. In particular, for the $ADE$ WZW models we have
$$ \hat B = C^X \otimes \bigl(C^{(l-1)}\bigr)^{-1}\ . \eqno(3.17)$$
Beyond giving a criterion for finding the $f_i$, the
physical meaning of the map $\beta$ as well of the structures
connected to it, \eg in algebraic $K$-theory, yet remains to
be clarified.
\bn\bn  {\bf 4. Further Conjectures And Outlook}   \bn
As already mentioned, the calculations of section 2 are by no means
restricted to the Virasoro minimal $c(2,q)$ models, but are
applicable to any rational CFT. This fact, and Kirillov's identities
for WZW models lead us to a general conjecture on the relation
between (effective) central charges $\cef$ and the Rogers' \dil:
\smallskip  \item{} Let
${\cal C}$ be the set of all possible values of $\cef$
occuring in non-trivial RCFT. ${\cal C}$ is additive
(since RCFT can be
tensorized), countable and supposed to be closed and well-ordered.
We conjecture that ${\cal C}$ is identical to the union ${\cal D}$
of the sets ${\cal D}^N$ of those rational numbers that can be
expressed in the form $\sum_{i=1}^N \L(x_i)/\L(1)$ where the
$x_i$ are algebraic numbers in the interval $(0,1)$. Moreover,
the $x_i$ producing $\cef$ of a certain RCFT should lie in the
field-extension of $\Q$ generated by the quantum dimensions of this
RCFT.
\sn
If this conjecture is true, it would shed a new
light on the task to classify all RCFT. Consider, e.g., ${\cal D}^1$,
which is believed to be the set $\{{1\over2},{2\over5},
{3\over5} \}$. Whereas ${1\over2} = \L({1\over2})/\L(1)$ corresponds
to the free Majorana fermion, the other two values should belong to
the $c(2,5)$ and $c(3,5)$ minimal models, where the only non-trivial
quantum dimension $\tau$ is given by the golden ratio $\tau^2=\tau+1$.
We have $\L(1/\tau^2)/\L(1) = {2\over5}$ for $c(2,5)$ (cf.\ section 2)
and $\L(1/\tau)/\L(1) = {3\over5}$ for $c(3,5)$.
This, in particular, implies that there
is just one possibility for a fusion ring with only one generator
except the identity, namely $\phi\times\phi=1+\phi$. From the mere
axioms of fusion rings, also $\phi\times\phi=1+n\phi$ with any natural
$n$ is allowed. Only the modular properties of RCFT enforce $n=1$,
which on the other hand comes out naturally from ${\cal D}^1$,
\ie from the \dil. Examples in ${\cal D}^{k-1}$ are provided by
eqs.\ (3.1) for $j=1$ and (3.5).        \sn
The dilogarithm function also appears in the calculation of volumina
of hyperbolic 3-manifolds (cf. \q{26,22}). More precisely, any
complete oriented hyperbolic 3-manifold $M$ of finite volume
(possibly with cusps) can be triangulated into
"ideal" tetrahedra $\Delta_1, \ldots, \Delta_n$,
each of which is specified by a complex number $z_{\nu}$.
The volume of a tetrahedron of this type
is given by $D(z_{\nu})$ and consequently
$ vol(M) = \sum_{\nu =1}^n D(z_\nu)$, where $D$ is the so-called
Bloch-Wigner function. This function up to some logarithmic
correction terms equals the imaginary part of the dilogarithm.
However, the function we have been looking
at is definitely the real part of the dilogarithm, so the quest is
for some geometrical interpretation of the latter. Indeed there is
one: According to work of Thurston, Meyerhoff, Neumann, Zagier and
Yoshida (cf. \q{25} and references therein) to $M$ one can associate
an invariant
$I(M)\in\C^{\times}$ having absolut value $\exp\{2\ vol(M)/\pi\}$,
whereas its argument is the Chern-Simons invariant $CS(M)$ of $M$.
Having corrected this
invariant by the length and torsion of certain geodesic loops,
one obtains a function which is complex analytic on a neighborhood
of the deformation space of $M$. In particular, for the hyperbolic
manifolds $M_{(p,q)}$ obtained by performing Dehn surgery of type
$(p,q)$ along
the figure-eight knot in $S^3$ ($p,q$ coprime, $|p|\leq 5$
if $|q|=1$) it was proven in \q{25} that
$$    12\ CS(M_{(p,q)}) + {3\over \pi}\ Tor\ \gamma
=- {6\over \pi^2} Re(L(z)+L(w)-L(1))\ (\mod\ 6)\, ,\eqno(4.1)$$
where the torsion $Tor$ is given by
$$ {3\over\pi}\ Tor\ \gamma = \bigl({6 r\over p}
        -{6\over\pi p} arg(z(1-z))\bigr)\ (\mod\ 6) \eqno(4.2)$$
and $r\in\{0,1,\ldots,|p|\}$ by $qr=1(\mod p)$;
$z$ and $w$ are complex numbers with positive real part which satisfy
$$ \eqalign{ &\log \ z + \log(1-z) + \log \ w + \log(1-w) = 0, \cr
&p\ \log(w(1-z)) + q\ \log(z^2(1-z)^2) = 2\pi i .\cr} \eqno(4.3)$$
The similarity to the Bethe Ansatz type equations (3.15) is obvious.
(4.3) indicates that the latter should be understood as equations
on the universal covering of $\C^{\times}$ -- which might be
responsible for the argument-corrections in (3.7). Of course
one could speculate that the torsion, on the other hand, is
connected to the corrections appearing in the conformal dimensions
(3.8). Since topological field theories related to RCFT's live
on 3-manifolds, a pairing of the respective invariants is natural,
including a prominent role for knots.   \sn
In this paper we have discussed a method which -- in addition to
a thorough exploitation of the \dil's functional equations --
might be sufficient to prove at least some of the conjectures
given here or previously. A deeper understanding, however, will
certainly involve the above-mentioned mathematical branches where
the \dil\ -- mysteriously enough -- shows up.
\sn
There are more concrete problems, though.
First, it would be nice if we could write at least one character of
any CFT in a sum-form as in the Andrews-Gordon identities
(see \q{19} for an interpretation of this in the language of affine
Lie algebras)
in order to generate \dil\ expressions for the central charge and
possibly for the conformal dimensions in any RCFT. Ways to achieve
this might be to study the annihilating ideals \q{6} of the
models and perhaps to
find path representations of the highest weight modules
\q{11,12}, or to introduce suitable filtrations on the space of fields.
However, as these sum-formulas need not be at all
unique, this could lead to further relations for the \dil\ function.
\sn
Another open problem is to establish a direct connection between the
\dil\ equations obtained from character \asy s to those involving the
fusion rules. Probably such a link would bring in the modular
properties of CFT and could also give some hints
concerning the rather opaque role of the "$\beta$-condition" in CFT.
\sn
Finally, the topics of integrable QFT and lattice models have only
been touched very briefly -- in agreement with the aim of this letter.
Nevertheless, we believe that the study of the \dil\ will reveal
new connections between these subjects and CFT. At least in
some cases CFT fusion rules appear as the (stationary)
"core" of the Thermodynamic Bethe Ansatz equations
("Baxterization of fusion rules"). We also expect that the
"canonical sum-form" of a character as propagated above will turn
out to be related to perturbations of CFT (cf.\ \q{15}).
Another interesting physical interpretation, however, has been
given very recently in \q{10} in terms of quasi-particle excitations
in quantum chains.
\sn\bn\bn
\vfill\eject
We would like to thank A.\ Goncharov, A.N.\ Kirillov, D.\ Zagier,
J.\ Kellendonk, M.\ R\"osgen and R.\ Varnhagen for various
discussions on related subjects. We are indebted to A.\ Kuniba for
pointing out a mistake in an earlier version of this paper.
\hfil\break  \noindent  A.R.\ is
supported by the Max-Planck-Institut f\"ur Mathematik, Bonn.
\bn\bn\leftline{{\bf References }} \bn
\halign{\hfil $ # $\phantom{xx}& \rm # \hfill  \cr
\q{1}&G.E.\ Andrews,{\sl q-series: their development and application
              in analysis, number}\cr
     &{\sl theory, combinatorics, physics, and computer
         algebra}, \cr
     &Conf.\ Board of the Math.\
      Sciences, Reg.\ Conf.\ Ser.\ in Math.\ {\bf66} (1886) \crns
\q{2}&V.V.\ Bazhanov, N.Yu.\ Reshetikhin, \intj {\bf4} (1989) 115\crns
\q{3}&A.\ Beilinson, A.\ Goncharov, V.\ Schechtman, A.\ Varchenko, \cr
     &in {\sl The Grothendieck Festschrift}, vol. 1, Birkh\"auser 1990,
        p.\ 135\crns
\q{4}&D.M.\ Bressoud, {\sl Analytic and combinatorical generalizations
         of the Rogers-} \cr
     &{\sl Rama\-nu\-jan
     identities}, Memoirs of the Am.\
         Math.\ Soc.\ No.\ 227, Vol.\ 24 (1980) \crns
\q{5}&R.\ Dijkgraaf, E.\ Verlinde, \np (Proc.\ Suppl.) {\bf5B}
         (1988) 87\crns
\q{6}&B.L.\ Feigin, T.\ Nakanishi, H.\ Ooguri, \intj
               {\bf 7} Suppl.\ {\bf 1A} (1992) 217 \crns
\q{7}&J.\ Fuchs, Int. J. Mod. Phys. {\bf B6} (1992), 1951 \crns
\q{8}&A.\ Goncharov, private communication \crns
\q{9}&V.G.\ Kac, {\sl Infinite Dimensional Lie Algebras},
        Cambridge 1990 \crns
\q{10}&R.\ Kedem, B.M.\ McCoy, {\sl Construction of modular branching
       functions from} \cr
      &{\sl Bethe's equations in the 3-state Potts chain},
       Stony Brook preprint ITP-SB-92-56 \crns
\q{11}&J.\ Kellendonk, A.\ Recknagel,
           {\sl Virasoro representations on fusion graphs},\cr
      &preprint BONN-HE-92-22, to appear in Phys.\ Lett.\ B \crns
\q{12}&J.\ Kellendonk, M.\ R\"osgen, R.\ Varnhagen, in preparation \crns
\q{13}&A.N.\ Kirillov, J.\ Sov.\ Math.\ {\bf47} (1989) 2450,
        (Zap. Nauch. Semin. LOMI {\bf 164} (1987) 121) \crns
\q{14}&A.N.\ Kirillov, N.Yu.\ Reshetikhin, J.\ Sov.\ Math.\ {\bf 52}
       (1989) 3156,\cr
       &(Zap. Nauch. Semin. LOMI {\bf 160} (1987) 211) \crns
\q{15}&T.R.\ Klassen, E.\ Melzer, \np {\bf338} (1990) 485, {\bf370}
        (1992) 511 \crns
\q{16}&A.\ Kl\"umper, P.A.\ Pearce, J.\ Stat.\ Phys.\ {\bf64}
         (1991) 13 \crns
\q{17}&A.\ Kuniba,  {\sl Thermodynamics of the $U_q(X^{(1)}_r)$
         Bethe Ansatz System} \cr
      &{\sl  with q a root of unity}, Canberra preprint SMS-098-91\crns
\q{18}&A.\ Kuniba, T.\ Nakanishi, {\sl Spectra in conformal field
          theories from the Rogers}\cr
      &{\sl dilogarithm}, preprint hep-th/9206034, SMS-042-92;
       {\sl Rogers \dil}\cr
      &{\sl  appearing in integrable systems}, Harvard
       preprint HUTP-92/A046 \crns
\q{19}&J.\ Lepowsky, R.\ Wilson, Proc.\ Natl.\ Acad.\ Sci.\ U.S A.
      {\bf78} (1981) 7254;  \cr
      &Invent.\ Math.\ {\bf77} (1984) 199, {\bf 79}, (1985) 417\crns
\q{20}&L.\ Lewin, {\sl Polylogarithms and associated functions},
        Elsevier North-Holland (1981) \crns
\q{21}&W.\ Nahm, Talk given at the Isaac Newton Institute, Cambridge,
           Sept.\ 1992; \cr
      &to appear in the proceedings of the NATO workshop {\sl
           Low-dimensional Topology}\cr
      &{\sl and Quantum Field Theory}\crns
\q{22}&W.D.\ Neumann, D.\ Zagier, Topology {\bf24} (1985) 307 \crns
\q{23}&F.\ Ravanini (1992), Phys.\ Lett.\ B{\bf282} (1992) 73; \cr
      &F.\ Ravanini, R.\ Tateo, A.\ Valleriani, {\sl Dynkin TBA's},
       Bologna preprint DFUB-92-11 \crns
\q{24}&B.\ Richmond, G.\ Szekeres,
   J.\ Austral.\ Math.\ Soc.\ Ser.\ A {\bf31} (1981) 362 \crns
\q{25}&T.\ Yoshida, Invent.\ Math.\ {\bf81} (1985) 473\crns
\q{26}&D.\ Zagier, {\sl The remarkable dilogarithm} (Number theory and
          related topics. \cr
      &Papers presented at the Ramanujan colloquium, Bombay 1988,
             TIFR)\crns
\q{27}&Al.B.\ Zamolodchikov, \np {\bf342} (1990) 695\crns
}\vfill\eject\end